\documentclass[pra,twocolumn,amsmath,amssymb]{revtex4-1} %

\usepackage{epsfig,amsmath}
\usepackage{subfigure}
\usepackage{graphicx}
\usepackage{dcolumn}
\usepackage{stmaryrd}
\usepackage{mathrsfs}
\usepackage{pifont}
\usepackage{amsthm}
\usepackage{amssymb}
\usepackage{bm}
\usepackage{latexsym}
\usepackage{hyperref}
\usepackage{color}

\newcommand{\la}{\langle}
\newcommand{\ra}{\rangle}

\newcommand{\ti}{\tilde}
\newcommand{\ga}{\gamma}
\newcommand{\Ga}{\Gamma}

\newcommand{\De}{\Delta}

\newcommand{\si}{\sigma}

\newcommand{\om}{\omega}
\newcommand{\Om}{\Omega}
\newcommand{\de}{\delta}

\newcommand{\non}{\nonumber}
\newcommand{\pa}{\partial}

\def\apl#1{{ Appl. \ Phys. \ Lett.} {\bf#1}}

\def\jpb#1{{ J.\ Phys.\ B} {\bf#1}}

\def\pra#1{{ Phys.\ Rev. A\/} {\bf#1}}
\def\prb#1{{ Phys.\ Rev. B\/} {\bf#1}}

\def\prl#1{{ Phys.\ Rev.\ Lett.} {\bf#1}}
\def\pr#1{{ Phys.\ Rev.} {\bf#1}}
\def\sci#1{{ Science} {\bf#1}}

\def\rmp#1{{ Rev. \ Mod. \ Phys.} {\bf#1}}
\def\nat#1{{ Nature} {\bf#1}}

\begin{document}

\title{Decoherence of a Driven Qubit}

\author{Jun Jing$^{1,2}$ \footnote{[Email address]: junjing@buffalo.edu}, Peihao Huang$^{1}$ \footnote{[Email address]: peihaohu@buffalo.edu}, and Xuedong Hu$^{1}$ \footnote{[Email address]: xhu@buffalo.edu}}

\affiliation{$^{1}$Department of Physics, University at Buffalo, SUNY, Buffalo, NY 14260, USA\\ $^{2}$Institute of Atomic and Molecular Physics, Jilin University, Chuangchun 130012, Jilin, China}

\date{\today}

\begin{abstract}
We study decoherence of a field-driven qubit in the presence of environmental noises. For a general qubit, we find that driving, whether on-resonance or off-resonance, alters the qubit decoherence rates (including dissipation and pure dephasing), allowing both blue and red sideband contributions from the reservoir. Depending on the noise spectral density, driving field detuning and driving field phase shift, the qubit decoherence rates could be either accelerated or reduced. We apply our general theory to the system of an electron spin qubit that is confined in a quantum dot and driven by an in-plane electric field. We analyze how spin relaxation induced by the electrical noise due to electron-phonon interaction varies as a function of driving frequency, driving magnitude, driving field phase shift and spin-orbit coupling strengths.
\end{abstract}


\maketitle

\section{Introduction}

Decoherence is a crucial issue in the studies of quantum information processing \cite{Nielsen} and the quantum-classical transition for physical systems \cite{Zurek_PhysToday}. As such, decoherence has been widely studied for many physical systems, ranging from atomic to solid state objects.

Among the multitude of decoherence studies, one branch focuses on the decoherence of a two-level system, or a qubit, because of its direct relevance to the quantum circuit model of a quantum computer \cite{Nielsen}. Such a study normally involves the coupling of an otherwise isolated and free qubit to an infinite reservoir (or bath) \cite{Meystre_QuantumOptics}, with the most famous being the spin-boson model \cite{Leggett_RMP88}. However, in a quantum information processor, there are inevitably many qubits, and they are often being driven by external fields, for example for single-qubit operations. Furthermore, selective single-qubit operations often involve shifting the frequency of a specific qubit relative to the others, then applying a global driving field that is on resonance with the selected qubit \cite{Kane_Nature99}. This approach for single-qubit operations means that many qubits would experience driving no matter whether they are being operated on or not. Therefore, investigating decoherence properties of a qubit while it is driven is a crucial step toward the building of a scalable quantum computer.

Studies of relaxation in an ensemble of weakly-driven spins go back to the early days of magnetic resonances \cite{Wangsness_PR53, Redfield_57, Ailion1, Ailion2, Abragam_book}. Over the past decade there have been both theoretical and experimental studies of decoherence of a driven qubit \cite{Geva_JCP95}, particularly on the decoherence of superconducting qubits, whether flux \cite{Smirnov, Han}, charge \cite{Ithier}, or phase qubits \cite{Kosugi, Yan}. These studies mainly focused on how a resonant driving field modifies the qubit decoherence rates, although there are also experimental evidences that off-resonant driving of a microwave resonator could lead to strong modifications of the flux-qubit Rabi frequency \cite{Gustavsson}. Other explorations include how to realize Landau-Zener transitions in superconducting persistent current qubit through longitudinal harmonic \cite{Oliver09} and biharmonic \cite{Oliver13} driving, how to infer the noise spectrum by studying the decay of Rabi oscillations of a flux qubit under strong driving \cite{Oliver14}, how decoherence affects the driving of a qubit (mostly spin qubits) \cite{RHuang, Dobrovitski, Koppens}, and how driving and decoherence affect tunneling through a double dot \cite{Barrett}. However, many interesting and important issues, such as effects of off-resonance driving by an external field, remain open, warranting further studies.

In this work, we develop a general theory on the decoherence of a driven qubit in a semi-classical noisy environment. In particular, we first calculate the qubit relaxation and dephasing rates in the reference frame that rotates at the frequency of the driving field, and find that the rates contain contributions from multiple frequencies of the noise spectrum, including both blue- and red-shifted sideband contributions. In the laboratory reference frame, the driving modifies the qubit decoherence (both relaxation/dissipation and pure dephasing) significantly as compared to a free qubit. In the case of resonant driving, we find the longitudinal relaxation rate in the lab frame is equivalent to the transverse relaxation rate in the rotating frame. In the case of off-resonance driving, the qubit decoherence is again influenced by both relaxation and dephasing in the rotating frame, though the calculation can only be carried through numerically in general. We then apply our theory to the case of a spin qubit that is driven electrically \cite{Tokura,Laird,You3} via the spin-orbit (SO) interaction, and under the influence of phonon noise \cite{Meier, Fabian98, Golovach_PRL04, Zutic_RMP04, Fabian06, Climente, Golovach_PRB08}. We find that the driven spin qubit undergoes both relaxation and dephasing, which is qualitatively different from the case of a free spin qubit, for which phonon noise leads to only relaxation at the lowest order of the SO interaction.

The rest of the paper is organized as follows. In Sec.~\ref{theory} we develop a general theory on the decoherence of a driven qubit influenced by a semi-classical noise based on the Bloch-Redfield method. In Sec.~\ref{effH}, we briefly summarize the theory on electrical driven spin resonance for a single spin qubit in a quantum dot. Details of the derivation of the effective spin Hamiltonian are given in Appendix \ref{derivation}. In Sec.~\ref{Rot}, we obtain the decoherence rates of this driven qubit under electron-phonon interaction in rotating frame. Secs.~\ref{res} and \ref{res_dephasing} are on the relaxation and pure dephasing properties (in lab frame) of a resonantly driven spin qubit, respectively. These discussions are further extended into the off-resonance cases in Secs.~\ref{off} and \ref{off_dephasing}. Lastly, we present further discussions and our conclusions in Secs.~\ref{Sec:discussion} and \ref{Conc}.

\section{General theory}\label{theory}

In this section we develop a general theory to treat the decoherence problem of a field-driven qubit. We first set up a model Hamiltonian, then simplify it by transforming into a reference frame that rotates with the driving field. This transformation allows us to examine the qubit decoherence in the rotating frame using the Bloch-Redfield approach, and then obtain qubit decoherence rates in the laboratory reference frame as well.

\subsection{Model Hamiltonian}

The effective Hamiltonian of a qubit driven by a classical field and under the influence of a noise could be written as (letting $\hbar = 1$)
\begin{equation}\label{Heff}
H_{\rm eff}=\frac{\om_Z}{2}\si_z+ \left(\frac{\Om}{2}e^{-i\nu t-i\phi}\si_++h.c. \right)+\sum_jn_j\si_j \ ,
\end{equation}
where $\om_Z$ is the qubit energy splitting, $\Om$, $\nu$ and $\phi$ are the strength, frequency and angle (between the driving field and the $x$-direction of the Bloch sphere)of the driving field, respectively, and $n_j$'s, $j=x,y,z$, are the three components of the noise experienced by the qubit.

In this study we assume that the noise is weak, $|n_j|\ll\om_Z, \Om, \nu$, so that Bloch equations \cite{Slichter_book} can be used to describe the dynamics generated by $H_{\rm eff}$. We also assume that different noise components are statistically independent, i.e. $\la n_j(t_1)n_k(t_2)\ra=0,  j\neq k$, and invariant under temporal translation, i.e. $\la n_j(t_1)n_j(t_2)\ra=W_jS(t_2-t_1)$, where $W_j$ is the noise strength along direction $j$, and $S(t)$ is the noise correlation function. The spectral information of the noise can be obtained from its Fourier transform: $S(t) = \frac{1}{2\pi} \int_{-\infty}^{\infty} d\om S(\om)e^{-i\om t}$, where $S(\om)$ is the spectral density function satisfying $S(\om)=S(-\om)$. We can express the noise spectral function along the $j$-direction as $S_{j}(\om)=W_jS(\om)$.

The effective Hamiltonian (\ref{Heff}) here also describes a spin-$1/2$ particle in a constant magnetic field (with Zeeman splitting $\om_Z$), driven by a transverse AC magnetic field of frequency $\nu$ and magnitude $\Omega$, and under the influence of a random magnetic noise in all three directions. Thus our results can be visualized in terms of a driven spin undergoing Rabi oscillation in the presence of a magnetic noise.

To calculate qubit dynamics governed by $H_{\rm eff}$, we need to first remove the time dependence introduced by the driving term. This can be done by transforming into a frame rotating at the driving field frequency. Specifically, we perform a canonical transformation $S_T^{(1)}$, with $S_T^{(1)}=i\frac{\nu}{2}\si_zt$, so that the Hamiltonian is transformed to  $H_{\rm eff}^{(1)}=e^{S_T^{(1)}}H_{\rm eff}e^{-S_T^{(1)}}+i\hbar\pa_tS_T^{(1)}$:
\begin{equation}\label{Heff1}
H_{\rm eff}^{(1)}= -\frac{\De}{2}\si_z +\frac{\Om}{2}\si_{x'}+w_t\si_{x'}+u_t\si_{y'}+n_z\si_z \ ,
\end{equation}
where $\De \equiv \nu-\om_Z$, $w_t={\rm Re}[z_te^{-i\phi}]$, $u_t={\rm Im}[z_te^{-i\phi}]$, and $z_t=(n_x+in_y)e^{-i\nu t}$. We have rotated the $xy$ axes to $x'y'$ so that $\si_{x'} \equiv \si_x \cos\phi + \si_y \sin\phi$ and $\si_{y'}\equiv-\si_x\sin\phi+\si_y\cos\phi$.  The spectral densities of the transformed noise components $w_t$ and $u_t$ are related to the spectral density $S(\omega)$ of the original noise $n_j$:
\begin{eqnarray}\label{Sw}
&&S_w(\om)=\frac{W_x\cos^2\phi+W_y\sin^2\phi}{2}
\ti{S}(\nu, \om)\,, \\ \label{Su}
&&S_u(\om)=\frac{W_x\sin^2\phi+W_y\cos^2\phi}{2}\ti{S}(\nu, \om)\,, \\ \label{Side}
&&\ti{S}(\nu, \om)\equiv S(\nu+\om)+S(\nu-\om)\,.
\end{eqnarray}

The transformed Hamiltonian $H_{\rm eff}^{(1)}$ describes a free spin in a tilted effective magnetic field (in the $x'z$ plane) under the influence of a modified magnetic noise. In particular, for the transformed noise components $w_t$ and $u_t$, their spectral densities at any frequency are an average of the red- and blue-shifted values from the original spectral density function $S(\omega)$.

We perform a further rotation around the $y'$-axis within the rotating reference frame so that the quantization axis $z'$ is along the total effective field. The transformation matrix is $S_T^{(2)}=i\frac{\theta}{2}\si_{y'}$, with $\tan \theta=-\Om/\De$. The resulting effective Hamiltonian takes the simple diagonalized form
\begin{eqnarray}
H_{\rm eff}^{(2)}&=& \frac{\om'}{2}\si_{z'} + (w_t \cos \theta - n_z \sin \theta)\si_{x'}  \nonumber \\
& & + u_t\si_{y'}+(w_t \sin \theta + n_z \cos \theta)\si_{z'} \ , \label{H2}
\end{eqnarray}
where $\om'=\Om\sin \theta-\De\cos \theta = \sqrt{\Omega^2 + \Delta^2}$. In this rotating frame, the original time-dependent driven-qubit problem becomes the problem of a free qubit (with a re-normalized energy splitting $\om'$) under the influence of a reshaped noise in all three directions.

\subsection{Decoherence in the Rotating Reference Frame}

We now calculate the decoherence rates for a qubit in the rotating frame using the effective Hamiltonian~(\ref{H2}). Our calculation is within the Bloch-Redfield equation framework, where the relaxation and pure dephasing rates are given by $1/T'_1 = S_{x'x'}(\om') + S_{y'y'}(\om')$, and $1/T'_\phi=S_{z'z'}(0)$, respectively. Here the noise spectral densities take the form
\begin{eqnarray}
S_{x'x'}(\om) & = & 2 \left[S_w(\om) \cos^2\theta + S_z(\om) \sin^2 \theta \right] \,, \nonumber\\
S_{y'y'}(\om) & = & 2S_u(\om) \,, \nonumber \\
S_{z'z'}(\om) & = & 2 \left[S_w(\om) \sin^2\theta + S_z(\om)\cos^2\theta \right] \,. \nonumber
\end{eqnarray}
Thus the qubit relaxation and dephasing rates in the rotating frame are
\begin{eqnarray}
\frac{1}{T'_1} & = & 2W_z \sin^2\theta S(\om')+ \left[ W_x(\cos^2\theta\cos^2\phi+\sin^2\phi) \right. \nonumber \\
& & \left. + W_y(\cos^2\theta\sin^2\phi+\cos^2\phi) \right] \ti{S}(\nu, \om')\,, \label{T1pri} \\
\frac{1}{T'_\phi} & = & 2\sin^2 \theta \left( W_x\cos^2\phi+W_y\sin^2\phi \right)S(\nu) \nonumber \\
& & + 2W_z \cos^2\theta \ S(0)\,. \label{Tphipri}
\end{eqnarray}

There are several notable features to the qubit decoherence rates in the rotating frame. The longitudinal relaxation rate is not only determined by the noise spectrum at the qubit energy splitting $\omega'$, but also at the side-band frequencies $\omega' \pm \nu$. Pure dephasing is determined not only by longitudinal noise ($W_z$) at zero frequency [$S(0)$], but also transverse noise ($W_x$ and $W_y$) at the driving frequency $\nu$. This additional contribution to pure dephasing is the consequence of the effective noise contained in $H_{\rm eff}^{(2)}$, where the longitudinal noise contains both the longitudinal and transverse components of the original noise ($n_x$, $n_y$ and $n_z$). Therefore, even if a free qubit in the laboratory reference frame does not experience pure dephasing (i.e. $W_z \equiv 0$), it does in the rotating frame when it is driven.

When the driving field is on resonance with the qubit, the expressions for the decoherence rates are simpler and more transparent. Assuming that the transverse noise is isotropic, $W_x = W_y$, the qubit relaxation and dephasing rates in the rotating frame are
\begin{eqnarray}
\left.\frac{1}{T'_1}\right|_{\rm res} & = & 2W_z S(\Omega)+ W_x \ti{S}(\omega_Z, \Omega) \,, \label{T1pri_res} \\
\left.\frac{1}{T'_\phi}\right|_{\rm res} & = & 2 W_x \ S(\omega_Z) \,. \label{Tphipri}
\end{eqnarray}
Not surprisingly, these rates are modified significantly compared to a free qubit, as the environmental noise in the rotating frame is altered from the lab frame.  Notice that the zero-frequency noise spectral density $S(0)$ is not present in either decoherence rates. In other words, in a fast rotating reference frame, low-frequency noise, even if strong, has a diminished effect on the qubit. For example, $1/f$ noise plays an important role in the dephasing of a charge qubit \cite{Petta04,Schreier} and a singlet-triplet qubit \cite{Dial_PRL13}. However, this decoherence channel would be significantly suppressed if the charge qubit is driven strongly and measurements can be done in the rotating frame \cite{Yan}.

The relaxation and dephasing rates obtained here represent decoherence in the rotating frame, and are meaningful if qubit dynamics in the rotating frame is accessible experimentally, such as the case in a superconducting qubit \cite{Yan}.  Recently, rotating frame magnetometry has also been demonstrated with a single Nitrogen-Vacancy center in diamond \cite{Loretz}. A qubit in the rotating-frame can be thought of as a qubit dressed by the driving field, and the decoherence rates presented in this section are properties of such a dressed qubit. On the other hand, many experimental measurements are on observable quantities in the laboratory reference frame. In these cases we need to rotate back to the lab frame (or switch back to the bare qubit) in order to quantify the effects of driving.  Below we discuss some general features of driven qubit decoherence in the lab frame.

\subsection{Relaxation in the Lab Frame}

We first focus on driven qubit relaxation in the laboratory frame. To have a consistent description, we set the qubit initial state at $\la\si_z(0)\ra=1$. In other words, it is in the excited eigenstate of $\sigma_z$ in the lab frame. The evolution of $\la\si_z(t)\ra$ can then be obtained as
\begin{eqnarray}\non
\la\si_z(t)\ra&=&\sin^2\theta \ e^{-t/T'_2}\cos\om't \\
\label{Siz} &+ & \cos \theta \left[\si'_\infty + \left( \cos \theta - \si'_\infty \right) e^{-t/T'_1} \right]\, ,
\end{eqnarray}
so that the longitudinal relaxation rate $1/T_1$ can be obtained numerically by setting the envelope of $\la\si_z(T_1)\ra$ at $\la\si_z(T_1)\ra = e^{-1}$. Here $1/T'_2 \equiv 1/(2T'_1)+1/T'_\phi$ is the transverse relaxation rate in the rotating frame, and $\si'_\infty$ is the asymptotic value in the long-time limit for $\la\si_z(t)\ra$ in the rotating frame, which is determined by the effective temperature of the modified noise. Note that usually $1/T_1$ is determined by both the relaxation rate $1/T'_1$ and the dephasing rate $1/T'_\phi$ from the rotating frame.

For a resonantly-driven qubit, its relaxation rate can be obtained analytically.  Specifically, when $\nu=\om_Z$, $\theta=\frac{\pi}{2}$ and $\om'=\Om$, so that $\la\si_z(t)\ra=e^{-t/T'_2}\cos\Om t$. Now the longitudinal relaxation rate $1/T_1$ in the lab frame is equivalent to the transverse relaxation rate $1/T'_2$ in the rotating frame:
\begin{eqnarray}\non
&& \left.\frac{1}{T_1}\right|_{\rm res} = \frac{1}{T'_2}=\frac{W_x\sin^2\phi+W_y\cos^2\phi}{2}\ti{S}(\om_Z, \Om) \\
&+&2\left( W_x\cos^2\phi+W_y\sin^2\phi \right)S(\om_Z) + W_z S(\Om)\,. \label{T_1}
\end{eqnarray}
Notice that the environmental noise contributes to qubit relaxation at multiple frequencies. In addition to the normal contribution at the qubit frequency $\omega_Z$, there are also sideband contributions at $\omega_Z \pm \Omega$, and a contribution at the Rabi frequency $\Omega$. These additional contributions are all consequences of driving.

Interestingly, when the qubit Rabi frequency is low, its relaxation rate $(1/T_1)|_{\rm res}(\Om \rightarrow 0)$ in Eq.~(\ref{T_1}) does not approach the relaxation rate for a free qubit, which is
\begin{equation}\label{nondr}
\left.\frac{1}{T_1}\right|_{\rm non-driven}=2(W_x+W_y)S(\om_Z) \,.
\end{equation}
The modification of the qubit relaxation due to resonant driving is
\begin{eqnarray}
&&\de\left(\frac{1}{T_1}\right)\equiv\left.\frac{1}{T_1}\right|_{\rm res}-\left.\frac{1}{T_1}\right|_{\rm non-driven} = W_z S(\Om) \nonumber \\
&& - \left(W_x\sin^2\phi+W_y\cos^2\phi\right) \left[2S(\om_Z)-\frac{1}{2}\ti{S}(\om_Z, \Om)\right]. \label{delta}
\end{eqnarray}
This modification arises because the external driving allows the qubit to sense the noisy environment in different frequency regions, while also redistributes the noise correlation strength along different directions. Mathematically, the (resonantly) driven and non-driven Hamiltonians are of different forms: one time-dependent, the other time-independent. To reach the non-driven limit from the driven Hamiltonian, one needs to first take $\nu \rightarrow 0$ to recover a stationary system Hamiltonian, and then let $\Omega \rightarrow 0$.

A previous study of resonantly driven tunneling in a double quantum dot found that in an Ohmic environment for a spin-boson model, $\de(1/T_1)$ is always smaller than zero \cite{Hanggi_ChemP04}. This ``coherent destruction of tunneling'' was identified as a phenomenon similar to motional narrowing or spin echo in spin resonance \cite{Slichter_book, Abragam_book, Zutic_RMP04}. In the current study, where we consider a generic noise, this regime of reduced relaxation is present as well. According to Eq.~(\ref{delta}), qubit relaxation is suppressed [$\de(1/T_1)<0$] when $W_z$ is relatively small (as compared to $W_x$ and/or $W_y$), and the noise spectral density is more or less flat [$S(\om_Z \pm \Om) \approx S(\om_Z)$]. For example, if the transverse noise is much stronger than the longitudinal noise, $W_x \sim W_y \gg W_z$, Eq.~(\ref{delta}) can be simplified to
\begin{equation}\label{twofold}
\de\left(\frac{1}{T_1}\right) \approx -\frac{1}{4} \left.\frac{1}{T_1}\right|_{\rm non-driven} \ .
\end{equation}
Here the relaxation rate for a driven qubit is reduced to $\sim 75\%$ of that in the non-driven case, and the result is independent of the properties of $S(\omega)$.

It is important to point out, however, that qubit relaxation is {\it not always suppressed} by driving. A simple example is a qubit in the presence of an isotropic white noise, for which $W_x=W_y=W_z$ and $S(\om)$ has no $\om$-dependence. In this case $\de(1/T_1)=0$, i.e. there is no difference between the relaxation rate of a driven and a non-driven qubit. For this special noise spectrum, the suppressed decoherence effect of the transverse noise is compensated by the enhanced effect of the longitudinal noise, and the net effect on relaxation vanishes.

In general, enhanced relaxation, i.e. $\de(1/T_1) > 0$, is also possible if the free qubit experiences only pure dephasing ($W_z \gg W_x, W_y$), or more generally if the qubit environment has some structures and/or anisotropy.  Consider the example when the noise has a Lorentzian spectral function,
\begin{equation}
S(\om)=\frac{1}{2\pi}\frac{\Ga\ga^2}{\ga^2+(\om-\om_c)^2},
\end{equation}
where $\ga$ is inversely proportional to the environment memory time (the smaller the $\ga$, the higher the spectral peak) and $\om_c$ corresponds to the peak frequency. In the special cases of $\om_Z \pm \Om = \om_c$, the regime $S(\om_Z)<\frac{1}{4}\ti{S}(\om_Z, \Om)$ could be accessible through tuning of $\ga$ or $\Omega$. With a further help from a non-zero longitudinal noise strength $W_z$, there could certainly exists a realistic parameter regime in which qubit relaxation is enhanced by driving.

We note here that our calculation of $1/T_1$ is based on the Bloch-Redfield equation. It is only valid for a Markovian or a near-Markovian environment, which corresponds to a relatively flat and smooth noise spectral function. To study a qubit in an environment with a sharply peaked noise spectral density, a non-Markovian treatment is required, and is beyond the scope of the current analysis.

\subsection{Pure Dephasing in the Lab Frame}

Information on pure dephasing is extracted from relaxation of the transverse components of the driven qubit. To allow a proper measurement of the transverse relaxation rate $1/T_2$ in the lab frame for the qubit, we prepare it initially at $\la\si_x(0)\ra=1$, i.e. in a $\sigma_x$ eigenstate. The dynamics of $\la\si_x(t)\ra$ under Hamiltonian (\ref{Heff}) can then be obtained as
\begin{eqnarray}\non
\la\si_x(t)\ra & = & \left[\si'_\infty \sin\theta + \left( \cos\phi \sin\theta - \si'_\infty \right) \sin\theta \ e^{-t/T_1'} \right.\\
\non & + & \left. S_\bot e^{-t/T_2'} \cos(\om' t + \psi) \cos\theta \right] \cos(\nu t+\phi) \\
\label{Six} & - & S_\bot e^{-t/T_2'} \sin(\om' t + \psi) \sin(\nu t+\phi)\,,
\end{eqnarray}
where $S_\bot \equiv \sqrt{\cos^2\theta \cos^2\phi + \sin^2\phi}$ and $\psi \equiv -\sin^{-1}\left(\sin\phi/\sqrt{\cos^2\theta \cos^2\phi + \sin^2\phi}\right)$. Clearly, the decay of $\la\si_x(t)\ra$ is governed by both $1/T_1'$ and $1/T_2'$, similar to the case of $\la\si_z(t)\ra$. The transverse polarization depends on the driving angle $\phi$ explicitly because the initial state is assumed to be polarized along the $x$-direction on the Bloch sphere. The presence of multiple sinusoidal functions in $\la\si_x(t)\ra$ leads to beatings and generally more complexities than in $\la\si_z(t)\ra$. In general the relaxation rate $1/T_2$ for the decay of $\la\si_x(t)\ra$ can only be determined numerically by setting the envelope of $\la\si_x(T_2)\ra$ at $e^{-1}$. The pure dephasing rate $1/T_\phi$ can then be obtained from $1/T_\phi=1/T_2-1/(2T_1)$.

Sideband effects on pure dephasing are difficult to extract from purely numerical solutions. However, on resonance and with special driving angles, pure dephasing rate could be obtained analytically. For example, with resonant driving and when $\phi=0$, the initial decay rate for $\la\si_x(t)\ra$ is $\sim 2/T_1'$ according to Eq.~(\ref{Six}) ($\si'_\infty$ is taken as $-1$, which is valid for many types of noises, such as $1/f$ noise and thermal noise), so that we can use $1/T_2=2/T_1'$ to represent the transverse relaxation rate. From Eqs.~(\ref{T1pri}), (\ref{Tphipri}) and (\ref{T_1}), we find
\begin{eqnarray}\label{phi0}
\left.\frac{1}{T_\phi}\right|_{{\rm res},\phi=0} & = & \frac{7}{4T_1'} - \frac{1}{2T_\phi'} \\ \non
& = & \frac{7}{2} W_z S(\Om) + \frac{7}{4} W_y \ti{S}(\om_Z, \Om) - W_x S(\om_Z) \,.
\end{eqnarray}
Recall that for a free qubit, the pure dephasing rate is determined by the noise spectrum at zero frequency:
\begin{equation}\label{nondrsx}
\left.\frac{1}{T_\phi}\right|_{\rm non-driven} = 2 W_z S(0) \,,
\end{equation}
thus $(1/T_\phi)|_{{\rm res},\phi=0}$ and $(1/T_\phi)|_{\rm non-driven}$ are determined by different parts of the noise spectrum, and are therefore unrelated to each other. Similar to the case in the rotating frame, non-vanishing pure dephasing could be generated by driving in the lab frame, even if a free qubit does not experience any pure dephasing.

In summary, from the analysis of both dissipation and pure dephasing of a driven qubit, we observe that (i) external driving dramatically modifies the decoherence rates ($1/T_1$ and $1/T_\phi$, and $1/T'_1$ and $1/T'_\phi$) in both the lab frame and the rotating frame through environmental noise redistribution and sideband contributions; (ii) in the case of resonant driving, $1/T_1$ in the lab frame is the same as $1/T_2'$ in the rotating frame; and (iii) driving can generate finite pure dephasing even if there is no pure dephasing for a free qubit.

\section{Application to an electrically driven spin qubit}

In this section we apply our general theory on the decoherence of a driven qubit to the analysis of an electron spin qubit in a quantum dot (QD) driven electrically and under the influence of phonon noise. While an electron spin can be driven with the traditional spin resonance technique using an AC magnetic field \cite{Slichter_book, Abragam_book}, a faster alternative can be achieved via electrical driving. This electrically driven spin resonance technique takes advantage of the finite spin-orbit (SO) interaction in a semiconductor (the so-called electric dipole spin resonance, or EDSR) \cite{Golovach_PRB08, Rashba_PRL03, Rashba_PRB08}, or the presence of an inhomogeneous magnetic field \cite{Tokura}. Here we will focus on using EDSR to drive an electron spin qubit.

In a semiconductor QD, the fastest single-spin decoherence mechanism at low temperatures and in a finite magnetic field is the pure dephasing coming from the hyperfine interaction between the electron spin and the nuclear spins of the host material. However, there are various ways by which this dephasing effect can be reduced \cite{Abragam_book, DD}, not to mention that in Si, the effects of the nuclear spins can be strongly suppressed through isotopic purification \cite{Zwanenburg_RMP13}. Beyond the nuclear spins, phonon noise through spin-orbit interaction generally constitutes the next most important decoherence channel for a spin qubit in a finite field \cite{Golovach_PRL04, Huang}, and will be the noise we study in this section.

In the following we first present the effective Hamiltonian for a spin qubit undergoing EDSR in the presence of phonon noise. We then analyze the obtained decoherence rates in the cases of on- and off-resonance driving, and discuss the implications of our results.

\subsection{Effective spin Hamiltonian}\label{effH}

The system we study is a single electron confined in a 2D quantum dot in the $xy$-plane. The growth direction $z$ has a much stronger confinement, thus we neglect the orbital dynamics along $z$. The driving electric field is applied in the $xy$-plane. The SO interaction contains both Rashba and Dresselhaus contributions \cite{Zutic_RMP04}. We perform a routine procedure to eliminate the SO interaction to the first order \cite{Golovach_PRB08, Stano05, Huang} and obtain the following effective spin Hamiltonian (A detailed derivation could be found in Appendix \ref{derivation}):
\begin{eqnarray}\label{Heff2}
H_{\rm eff} & \approx & \frac{1}{2}g\mu_B(\vec{B}_0+\vec{B}_e)\cdot\vec{\si}, \\
\non \vec{B}_e & = & \frac{2e}{g\mu_B\om_d^2}[\beta_-\dot{E}_y, \beta_+\dot{E}_x, 0] \,.
\end{eqnarray}
For simplicity we have chosen a perpendicular applied magnetic field $\vec{B}_0 = B_z \vec{z}$ along the $z$ direction, and $\vec{B}_e$ is the effective magnetic field from the electrical driving field and the electrical noise through SO interaction. In addition, $\hbar \omega_d$ is the dot confinement energy, $\mu_B$ is the Bohr magneton, and $\beta_{\pm} \equiv \beta \pm \alpha$ are the SO interaction strengths, with $\alpha$ and $\beta$ the Rashba and Dresselhaus interaction strengths, respectively. Notice that here both driving and noise are along transverse directions. We can apply our general theory quite straightforwardly in this case, with $n_z=0$ and $\hbar \om_Z = g\mu_BB_z$, and the driving field and the noise satisfying the following relationships:
\begin{eqnarray}\non
\frac{\Om\cos(\nu t+\phi)}{2}&=&\frac{e\dot{E}_y^c}{\hbar\om_d^2}\beta_-, \quad
n_x=\frac{e\dot{E}_y^f}{\hbar\om_d^2}\beta_-, \\ \label{Et}
\frac{\Om\sin(\nu t+\phi)}{2}&=&\frac{e\dot{E}_x^c}{\hbar\om_d^2}\beta_+, \quad
n_y=\frac{e\dot{E}_x^f}{\hbar\om_d^2}\beta_+ \,.
\end{eqnarray}
Notice that the driving electric field obtained here is elliptically polarized when $\beta_- \neq 0$, i.e. $\alpha\neq\beta$. When $\beta_- = 0$ the driving field would be linearly polarized along $x$ axis, as we discuss in the Appendix. The ``driving angle'' $\phi$ in the general theory is a phase shift for the elliptically-polarized electric field. It gives the initial field direction, which affects the subsequent spin dynamics and how the spin senses the phonon reservoir through the SO interaction. The driving strength $\Omega$ from EDSR could be estimated as
\begin{equation}
\Om \approx 2\frac{(\beta+\alpha)e|E_x| \om_Z}{\hbar\om_d^2} \,,
\label{eq:Omega}
\end{equation}
which is proportional to the SO coupling strength, the magnitude of the driving electrical field, and the Zeeman splitting $\om_Z$, and inversely proportional to the square of the confinement energy $\omega_d$. With $\hbar\om_d \sim 1$ meV \cite{Hanson_RMP07}, $|E_x|\sim 4000$ V/m \cite{Nowack}, and $\beta\approx1000$m/s (for GaAs), we estimate that $\Omega/\omega_Z \sim 10^{-2}$. In other words, the driving field in the existing experiments is relatively weak, within the applicable regime of our theory above.

\subsection{Spin Decoherence in the Rotating Frame}\label{Rot}

We first calculate the relaxation and pure dephasing rates for the electron spin in the rotating frame. These rates are directly relevant if experimental measurements can be done in the rotating frame, like what have been done in superconducting qubits \cite{Yan} and NV centers \cite{Loretz}. They are also crucial in calculating lab-frame decoherence rates. From Eqs.~(\ref{T1pri}), (\ref{Tphipri}), and (\ref{Heff2}), we obtain
\begin{eqnarray}
\frac{1}{T'_1} & = & \frac{e^2}{\hbar^2\om_d^4} \Big[(\beta_-^2\sin^2\phi+\beta_+^2\cos^2\phi) \nonumber \\
& & + (\beta_-^2\cos^2\phi+\beta_+^2\sin^2\phi)\frac{\De^2}{\De^2+\Om^2}\Big] \ti{S}(\nu,\Om), \label{T1pri2} \\
\frac{1}{T'_\phi} & = & \frac{2e^2}{\hbar^2\om_d^4}(\beta_-^2\cos^2\phi+\beta_+^2\sin^2\phi) \frac{\Om^2}{\De^2+\Om^2}S(\nu), \label{Tphipri2}
\end{eqnarray}
where $\ti{S}(\nu,\Om)$ is defined in Eq.~(\ref{Side}). The noise electric field considered here is from the piezoelectric electron-phonon interaction, which is important in GaAs and InAs quantum dots \cite{Yu,Borhani}. The corresponding spectral function, including both longitudinal acoustic and transverse acoustic phonon branches, is $S(\om)=\hbar e^2_{14}|\om|^5/15\pi^2\rho c^5$. This spectral function is obtained at low temperature within dipole approximation, and with the assumption of an isotropic linear dispersion relation for the phonons. Here $e_{14}$ is an elasticity tensor component, $\rho$ is the mass density, and $c$ is the speed of sound in the substrate material for the QD. For larger applied magnetic fields, the deformation potential electron-phonon interaction may become more important, though the general results will be similar to what we obtain here.

For more concrete expressions of $1/T'_1$ and $1/T'_\phi$ in EDSR, we set $R \equiv \Om/\om_Z$ and $\delta \equiv \De/\om_Z$, which are dimensionless driving strength and detuning, respectively. We also introduce $r \equiv \alpha/\beta$ to represent the relative strength of the Rashba SO coupling, and we take the Dresselhaus strength $\beta$ as intrinsic and fixed. After introducing the phonon spectral density, Eqs.~(\ref{T1pri2}) and (\ref{Tphipri2}) now take the form
\begin{eqnarray}\label{T1De}
&& \frac{1}{T_1'} = \frac{e^2e_{14}^2\beta^2\om_Z^5}{15\pi^2\hbar\rho c^5\om_d^4}(1+\delta)^5 F\left(\frac{R}{1+\delta}\right) \\
&\times& \left[ \left( 1 + r^2 -2r \cos2\phi \right) \frac{\delta^2}{R^2+\delta^2} + \left(1 + r^2 + 2r \cos 2\phi\right) \right], \non  \\
\label{taylor} && F(x)\equiv (1-x)^5+(1+x)^5 \,,
\end{eqnarray}
and
\begin{equation}
\frac{1}{T'_\phi} = \frac{e^2e_{14}^2 \beta^2 \om_Z^5}{15\pi^2\hbar\rho c^5 \om_d^4} \frac{2R^2 (1 + \delta)^5}{R^2 + \delta^2} \left( 1 + r^2 - 2r \cos2\phi \right) \,.  \label{TphiDe}
\end{equation}
Both decoherence rates are factored into a dimensionless part determined by driving, detuning, and relative Rashba strength, and a common prefactor that is determined by the Dresselhaus strength $\beta$, the qubit energy splitting $\omega_Z$, and the dot confinement energy $\omega_d$.

\begin{figure}[htbp]
\centering
\includegraphics[width=3.2in]{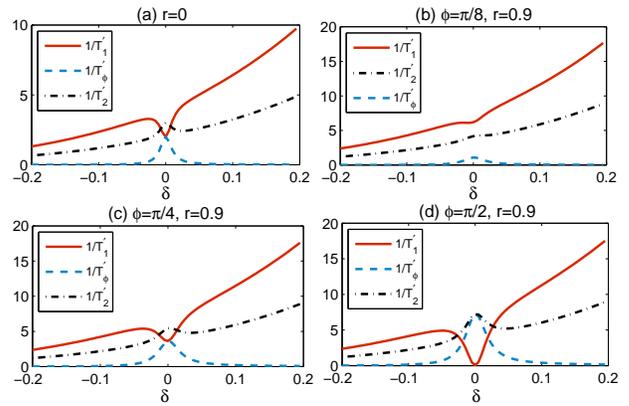}
\caption{(Color online) Dimension-less [without multiplying the common pre-factor in the expressions of decoherence times in Eqs.~(\ref{T1De}) and (\ref{TphiDe})] rotating-frame decoherence rates $1/T_1'$, $1/T_\phi'$, and $1/T_2'$ as functions of $\de$, with different combinations of $\phi$ and $r$. Note that when $r=0$, none of these rates depends on $\phi$.}\label{T1T2Tphi}
\end{figure}

To have a qualitative comparison of the different decoherence rates in the rotating frame, in Fig.~\ref{T1T2Tphi} we plot the relative values of $1/T_1'$, $1/T_\phi'$, and $1/T_2'$ as functions of the dimensionless detuning $\delta$, with different combinations of SO strength ratio $r$ and driving field phase shift $\phi$. There are roughly two regimes for all the curves in Fig.~\ref{T1T2Tphi}: the large-detuning regime, when $|\De| \gg \Om$ (or $|\delta| \gg R$); and the near-resonance regime, when $|\De| \lesssim \Om$ ($|\delta| \lesssim R$). Below we discuss these two regimes in more details.

For large detunings, when $|\De| \gg \Om$, i.e. $|\delta| \gg R$, Eqs.~(\ref{T1De}) and (\ref{TphiDe}) can be further simplified to
\begin{eqnarray}
\left.\frac{1}{T_1'}\right|_{|\delta| \gg R} & \approx & \frac{4 e^2e_{14}^2\beta^2\om_Z^5}{15\pi^2\hbar\rho c^5\om_d^4}(1 + \delta)^5 \left(1 + r^2 \right), \label{T1DeL} \\
\left.\frac{1}{T'_\phi}\right|_{|\delta| \gg  R} & = & \frac{2 e^2e_{14}^2 \beta^2 \om_Z^5}{15\pi^2\hbar\rho c^5 \om_d^4} \frac{R^2}{\delta^2} (1 + \delta)^5 \nonumber \\
& & \times \left( 1 + r^2 - 2r \cos2\phi \right) \,. \label{TphiDeL}
\end{eqnarray}
In the large-detuning limit, $1/T_1'$ does not depend on $\phi$. This is clearly demonstrated by the solid curves in Fig.~\ref{T1T2Tphi}(b) through ~\ref{T1T2Tphi}(d), all of which have the same value at the same detuning at the large detuning limit. $1/T_1'$ also has a quadratic dependence on $r$, which is illustrated by the solid curves in Figs.~\ref{T1T2Tphi}(a) and \ref{T1T2Tphi}(b), with $1+r^2 = 1$ and $\sim 2$, respectively. The general increasing trend for $1/T'_1$ in Fig.~\ref{T1T2Tphi} comes from the $(1+\delta)^5$ dependence, which is in turn from the phonon noise spectral density. For $1/T'_\phi$, on the other hand, the spectral-density dependence is largely dominated by the $R^2/\delta^2$ dependence in our current (and experimentally typical) parameter regime.

Near resonance, $1/T'_\phi$ clearly has a Lorentzian maximum at $\delta=0$ from the $R^2/(R^2 + \delta^2)$ dependence. Physically, at resonance and in the rotating frame, the qubit is precessing around the direction of the driving field, which is transverse to the quantization axis. The phonon noise shows up now as a longitudinal noise for the qubit in the rotating frame, therefore it can cause dephasing. Away from resonance, the driving does not cause much change in the state of the qubit, so that phonon noise is still a transverse noise and cannot cause dephasing.

The relaxation rate $1/T_1'$ in the rotating frame generally has a local minimum near resonance, as the part of the contributions that is proportional to $\delta^2/(R^2 + \delta^2)$ is suppressed when $|\delta| \ll R$. The transverse relaxation rate $1/T'_2$ thus features a competition between the opposing trends near resonance for $1/T_1'$ and $1/T'_\phi$, though Fig.~\ref{T1T2Tphi} shows that $1/T'_2$ usually has a slight bump in the near-resonance regime.

As we discussed in Sec.~\ref{theory}, our calculation of relaxation rates relies on the weak noise assumption, which requires that $|n_j|\ll\Om, j=x,y,z$. For GaAs, the noise electric field due to electron-phonon interaction is roughly $|\vec{E}^f| \lesssim 10$ V/m, while the driving field $|\vec{E}^c|$ could be up to $4000$ V/m in EDSR experiments on a single electron spin \cite{Nowack}. Therefore for EDSR in GaAs the weak noise assumption is valid as long as the external field and the driving field are not too small, and our general theory is applicable.

With knowledge of decoherence rates in the rotating frame, we are now ready to examine the decoherence properties of an electron spin qubit in the laboratory frame. In the following subsections we will discuss qubit relaxation and pure dephasing under resonant and off-resonance driving.

\subsection{Relaxation under Resonant Driving}\label{res}

When the driving field is on resonance with the electron spin Zeeman splitting, $\nu = \omega_Z$, i.e., $\de=0$, the longitudinal relaxation rate for the electron spin qubit in the lab frame can be expressed as the sum of a Zeeman contribution and a sideband contribution:
\begin{eqnarray}\label{T1lf}
\left.\frac{1}{T_1}\right|_{\rm res}&=&\left.\frac{1}{T_1}\right|_{\rm Zeeman}+\left.\frac{1}{T_1}\right|_{\rm sideband}, \\ \non
\left.\frac{1}{T_1}\right|_{\rm Zeeman}&=&\frac{2e^2}{\hbar^2\om_d^4}(\beta_-^2\cos^2\phi+\beta_+^2\sin^2\phi)S(\om_Z)
\\ &=& \frac{2e^2 \beta^2}{\hbar^2\om_d^4} \left( 1 + r^2 - 2r \cos 2\phi \right) S(\omega_Z) \,, \nonumber \\ \non
\left.\frac{1}{T_1}\right|_{\rm sideband}&=&\frac{e^2F(R)}{2\hbar^2\om_d^4}(\beta_-^2\sin^2\phi+\beta_+^2\cos^2\phi)S(\om_Z)
\\ &=& \frac{e^2 \beta^2 F(R)}{2\hbar^2\om_d^4} \left( 1 + r^2 + 2r \cos 2\phi \right) S(\omega_Z) \,, \nonumber
\end{eqnarray}
where the $F$ function is defined in Eq.~(\ref{taylor}).  We can normalize this driven qubit relaxation rate with respect to the free qubit rate given by
\begin{equation}\label{nd}
\left.\frac{1}{T_1}\right|_{\rm non-driven}=\frac{4e^2 \beta^2}{\hbar^2\om_d^4} (1 + r^2) S(\om_Z)\,.
\end{equation}
The resulting normalized relaxation rate is
\begin{eqnarray}\nonumber
\frac{\left( 1/T_1 \right)_{\rm res}}{\left( 1/T_1 \right)_{\rm non-driven}}
&=& \frac{1}{2} \left(1 - \frac{2r}{1+r^2} \cos2\phi\right)\\ &+& \frac{F(R)}{8} \left(1 + \frac{2r}{1+r^2} \cos2\phi\right) \,,
\end{eqnarray}
which is a function of the dimensionless driving strength $R$, SO strength ratio $r$, and the driving field phase shift $\phi$.

\begin{figure}[htbp]
\centering
\includegraphics[width=3.2in]{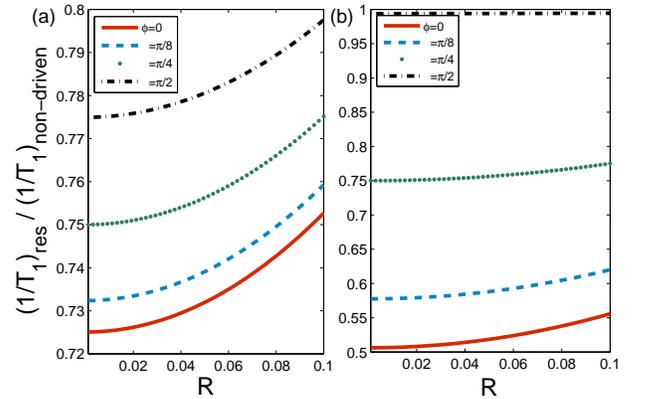}
\caption{(Color online) The ratio of on-resonance relaxation rate $(1/T_1)|_{\rm res}$ and non-driven relaxation rate $(1/T_1)|_{\rm non-driven}$ of electron spin as a function of $R$, the ratio of driving electrical field strength and Zeeman frequency, with different angles of driving electrical field. We choose (a) $r=0.05$ and (b) $r=0.8$. $B_z$, $\hbar\om_d$,  $|E_x|$ and $|E_y|$ are supposed to be tuned to realize $R$ to be in the range of $[10^{-3}, 10^{-1}]$.}\label{T1Om}
\end{figure}

In Fig.~\ref{T1Om} we present the dependence of the normalized spin relaxation rate on the driving strength $R$, which corresponds to Rabi frequency since the driving is on-resonance, at different $\phi$. There are several interesting features to the results in this figure.  First, the dependence on $R = \Omega/\omega_Z$ is quite weak.  When $R$ increases from $0.001$ to $0.1$, the relaxation rate increases at most about $10\%$. Second, the relaxation rate generally does not go back to the free qubit rate even when $\Omega$ is very small. Third, the driving field phase shift dependence is much more prominent in panel (b), when the Rashba and Dresselhaus SO coupling strength are similar ($r = 0.8$), compared to panel (a), when Dresselhaus coupling is dominant ($r = 0.05$).  Below we examine these features more closely.

The $R$-dependence of the relaxation rate comes completely from the sideband contribution, which is proportional to $F(R)$. As we discussed in the previous subsection, in the case of EDSR, $R$ tends to be small, up to about $0.01$ for GaAs with current technology \cite{Nowack, Petta}. Thus we can expand $F$ and obtain $F(R) \approx 2 + 20 R^2$. Even when $R = 0.1$, the correction to the value of $F$ is still only $10\%$, which means that the change to the overall relaxation rate due to a realistic finite $R$ is at most $10\%$. Indeed among all the curves presented in Fig.~\ref{T1Om}, only the $\phi = 0$ curve has a close-to-$10\%$ increase, because in this case the Zeeman contribution to the relaxation rate is strongly suppressed.

In Fig.~\ref{T1Om} the lower limit for the value of $R$ is $0.001$, not $0$. To maintain the validity of our weak noise assumption, the lower-bound of the driving field strength is $\Omega \gg 1/T_1$. Therefore, the small-$R$ data presented in Fig.~\ref{T1Om} should only be used as a benchmark for comparison with the higher-$R$ results, but is not the asymptotic value of the relaxation rate as $R \rightarrow 0$.  It is thus not such a surprise that the ratio $(1/T_1)|_{\rm res}/(1/T_1)|_{\rm non-driven}$ does not go to $1$ in general when $R$ is small. For the calculated $(1/T_1)|_{\rm res}$ to approach $(1/T_1)|_{\rm non-driven}$, one needs to take the limit $\nu \rightarrow 0$ first, and then let $\Omega \rightarrow 0$, as we have discussed in the previous section.

\begin{figure}[htbp]
\centering
\includegraphics[width=3.2in]{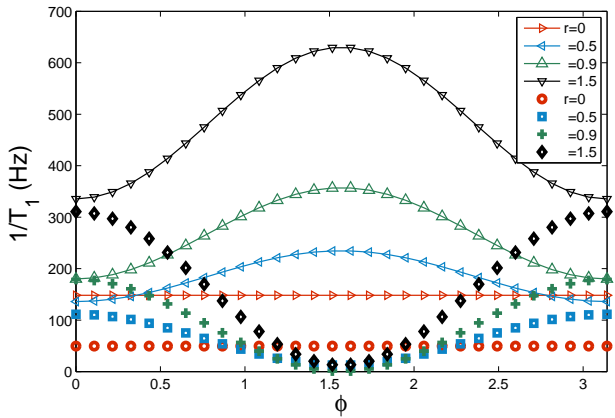}
\caption{(Color online) On-resonance relaxation rate $1/T_1$ (solid lines with symbols) of electron spin as a function of the ratio $r$ and the angle (phase shift) of driving electrical field in GaAs QDs, where we suppose $\beta=1000$m/s. The curves depicted only by symbols imply $(1/T_1)|_{\rm sideband}$, i.e., the term with $F(R)$ defined in Eq.~(\ref{taylor}). We choose $B_z=1$T, $\hbar\om_d=1$meV and $|E_x|=4000$V/m. }\label{T1b}
\end{figure}

$(1/T_1)_{\rm res}$ is a sinusoidal function (more specifically, $\cos 2\phi$) of the driving field phase shift $\phi$, as indicated in Eqs.~(\ref{T1lf}) and ~(\ref{nd}).  Furthermore, this $\phi$-dependence is the most prominent when $r \sim 1$, and is suppressed for $r \gg 1$ or $\ll 1$.  Thus we see the more dramatic $\phi$ dependence in panel (b) as compared to panel (a) in Fig.~\ref{T1Om}. For a more careful examination of the $\phi$-dependence, in Fig.~\ref{T1b} we plot the spin relaxation rate $(1/T_1)_{\rm res}$ as a function of $\phi$ with various values of $r$, which is fixed by the fabrication process for the quantum dot. The sideband contribution to $(1/T_1)_{\rm res}$ is indicated by curves without linking lines. By Eqs.~(\ref{T1lf}), $(1/T_1)_{\rm res}$ relies on $\cos(2\phi)$, so that it is symmetrical with respect to $\phi=\pi/2$. When $r=0$ (no Rashba SO coupling) or $r \rightarrow \infty$ (no Dresselhaus SO coupling), the relaxation rate and the sideband contribution are $\phi$-independent, as we have discussed above. When $r$ is finite, $(1/T_1)_{\rm res}$ reaches its maximum at $\phi=\pi/2$, while the sideband contribution is dominant when $\phi$ approaches $0$ or $\pi$. From the perspective of minimizing relaxation while driving, clearly a driving field phase shift near $0$ is preferable.

\begin{figure}[thbp]
\centering
\includegraphics[width=3.2in]{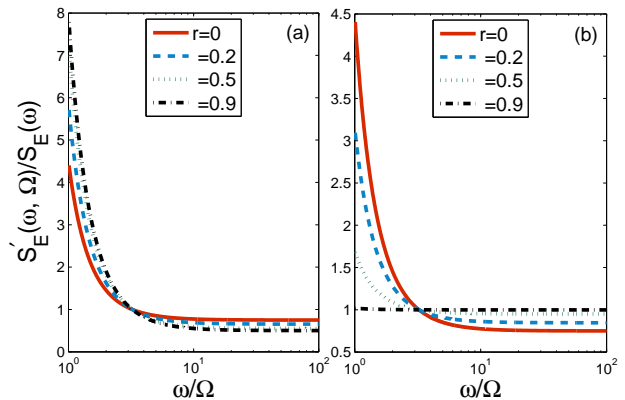}
\caption{(Color online) Modification of the piezoelectric electron-phonon spectrums for GaAs QD in unit of driven strength $\Om$ in Rabi oscillation. We choose driving field along the angle (a) $\phi=0$ and (b) $\phi=\pi/2$.}
\label{Spec}
\end{figure}

A driven qubit can be thought of as a free qubit in a modified environment $S'_E(\om, \Om)$, which can in turn be compared with the unmodified $S_E(\om)$. From Eqs.~(\ref{T1lf}) and (\ref{nd}), we find
\begin{eqnarray}\non
\frac{S'_E(\om, \Om)}{S_E(\om)} & = & \frac{1 + r^2 - 2r\cos2\phi}{2\left(1 + r^2\right)} \\
\non & +& \frac{1 + r^2 + 2r \cos2\phi}{8(1 + r^2)} F\left(\frac{\Om}{\om}\right) \,.
\end{eqnarray}
An interesting limit is when $\omega \gg \Omega$ (the reservoir is at the high frequency limit), so that $F(\Omega/\omega) \approx 2$. The above ratio is then simplified to
\begin{equation}
\label{SE}
\frac{S'_E(\om, \Om)}{S_E(\om)} \approx \frac{3}{4} -\frac{1}{2} \frac{r}{1+r^2} \cos 2\phi \,.
\end{equation}
In this domain $S'_E/S_E$ does not depend on $\omega$ at all, so that the rescaling of the environmental spectral density is uniform. This is clearly demonstrated in Fig.~\ref{Spec}, where we plot $S'_E(\om, \Om)/S_E(\om)$ at $\phi = 0$ and $\phi = \pi/2$ and with several different $r$. In the high-frequency domain for both panels in Fig.~\ref{Spec}, the re-scaling of the noise spectrum is independent of $\omega$. Practically, $\omega_Z$ falls in the high-frequency domain since $\Om$ is about $0.001\om_Z \sim 0.1 \om_Z$ in current GaAs QD experiments. The re-scale coefficient reaches its minimal value of $\frac{1}{2}$ when $r \sim 1$ and $\phi=0$, and its maximal value of $1$ when $r \sim 1$ and $\phi=\pi/2$.

In the low-frequency domain $\om/\Om\in[10^0,10^1]$, the driving generates a much more pronounced effect on the re-scaling coefficient, which is also quite sensitive to $r$. Note that while within EDSR the system generally falls into the high-frequency regime, our general theory on the driven qubit does not require that $\Omega$ is small compared to $\omega_Z$. As such there could be situations where the low-frequency end of the environment could be sensed by the qubit.

In short, the electron spin relaxation is significantly modified when it is driven resonantly, although the dependence on the driving strength is quite weak.

\subsection{Pure Dephasing under Resonant Driving}\label{res_dephasing}

For a free spin qubit, electron-phonon interaction does not cause pure dephasing \cite{Golovach_PRL04}. However, as we have discussed above, driving modifies the environment that the spin qubit experiences, so that pure dephasing due to electron-phonon interaction is non-vanishing for a driven spin qubit.

Generally, the pure dephasing rate $1/T_\phi$ can only be obtained numerically, although there do exist cases when analytical results can be found. For example, when the driving is on resonance with the spin qubit and the driving phase shift is $\phi=0$, the pure dephasing rate is
\begin{equation}
\left.\frac{1}{T_\phi}\right|_{{\rm res}, \phi = 0} = \frac{e^2 \beta^2}{\hbar^2\om_d^4} \left[\frac{7}{4} F(R) (1+r)^2 - (1-r)^2 \right] S(\om_Z),
\end{equation}
which is determined by the noise spectral density at the Zeeman splitting $\om_Z$ and the sideband frequencies $\om_Z \pm \Om$, instead of the zero-frequency limit for a free qubit.

\begin{figure}[htbp]
\centering
\includegraphics[width=3.2in]{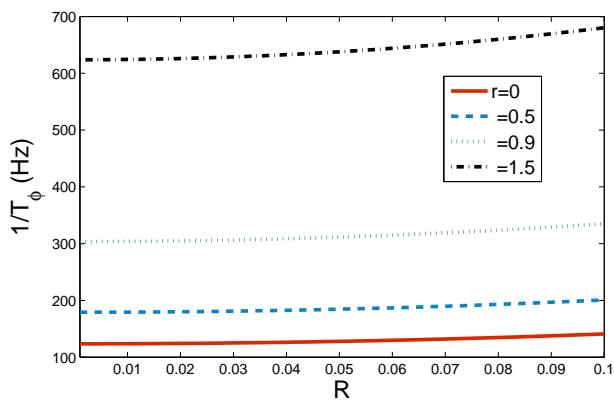}
\caption{(Color online) On-resonance pure dephasing rate $1/T_\phi$ of electron spin as a function of driving field amplitude $R$ at a few different ratios $r$ in GaAs QDs, where we have taken $\phi=0$ and $\beta=1000$m/s. $B_z$, $\hbar\om_d$, $|E_x|$ and $|E_y|$ are tuned such that $R$ is in the range of $[10^{-3}, 10^{-1}]$.}\label{Tphi0}
\end{figure}

We plot the driving strength ($R$) dependence of $(1/T_{\phi})_{{\rm res}, \phi = 0}$ in Fig.~\ref{Tphi0} with various SO ratio $r$. In the practically reasonable regime of weak driving, with $0.001\leq R\leq 0.1$ for EDSR, $F(R) \approx 2 + 10R^2 \sim 2$, so that $(1/T_\phi)_{{\rm res}, \phi = 0}$ is not sensitive to the driving strength $R$, just like $(1/T_1)_{\rm res}$. In this limit the pure dephasing rate is proportional to $5/2 + 9r + 5r^2/2$, which increases monotonically with $r$, with a larger $r$ indicating a larger overall strength of the SO interaction. Comparing Fig.~\ref{T1b} and Fig.~\ref{Tphi0}, we observe that $(1/T_\phi)_{{\rm res}, \phi = 0}$ is much larger than $(1/T_1)_{\rm res}$ when $\phi = 0$ and $r \rightarrow 1$, which means that for this parameter combination pure dephasing plays a more important role in spin decoherence than relaxation. Away from this particular parameter combination, the magnitudes of pure dephasing and relaxation are in the same order. We can thus conclude that for a resonantly driven spin qubit, pure dephasing is as important a decoherence channel as relaxation when electron-phonon interaction is considered.

\subsection{Relaxation under Off-Resonance Driving}\label{off}

When an off-resonance AC field (magnetic or electric) is applied to a spin qubit, the coherent evolution of the qubit state is a rotation along a tilted axis in the rotating frame, determined by $H= -\frac{\Delta}{2} \sigma_z + \frac{\Omega}{2}\sigma_{x'}$ [See Eq.~(\ref{Heff1})]. As field detuning increases, the rotation axis for the spin approaches the $z$-axis (which is the direction of the applied DC field), so that the spin evolves as if it is not driven.  However, when considering spin decoherence, the off-resonance driving field does have some significant effects, as demonstrated by Eqs.~(\ref{T1pri2}) and (\ref{Tphipri2}) for EDSR. In this subsection, we examine in more detail how the off-resonance driving field affects the longitudinal relaxation of a spin qubit in the lab frame.

The lab-frame relaxation rate $1/T_1$ is in general evaluated numerically using Eq.~(\ref{Siz}) by setting $\la\si_z(T_1)\ra=1/e$. We neglect the fast oscillation in Eq.~(\ref{Siz}) and use only the envelope for our calculation of $1/T_1$, such that
\begin{equation}\label{nonr}
\frac{R^2}{R^2 + \de^2} e^{-T_1/T'_2} + \frac{\de^2}{R^2 + \de^2}e^{-T_1/T'_1} = \frac{1}{e}.
\end{equation}
$1/T_1$ thus always falls between $1/T'_1$ and $1/T'_2 \equiv1/(2T'_1)+1/T'_\phi$.

\begin{figure}
\centering
\subfigure{\label{T1nonab:ab}
\includegraphics[width=3.2in]{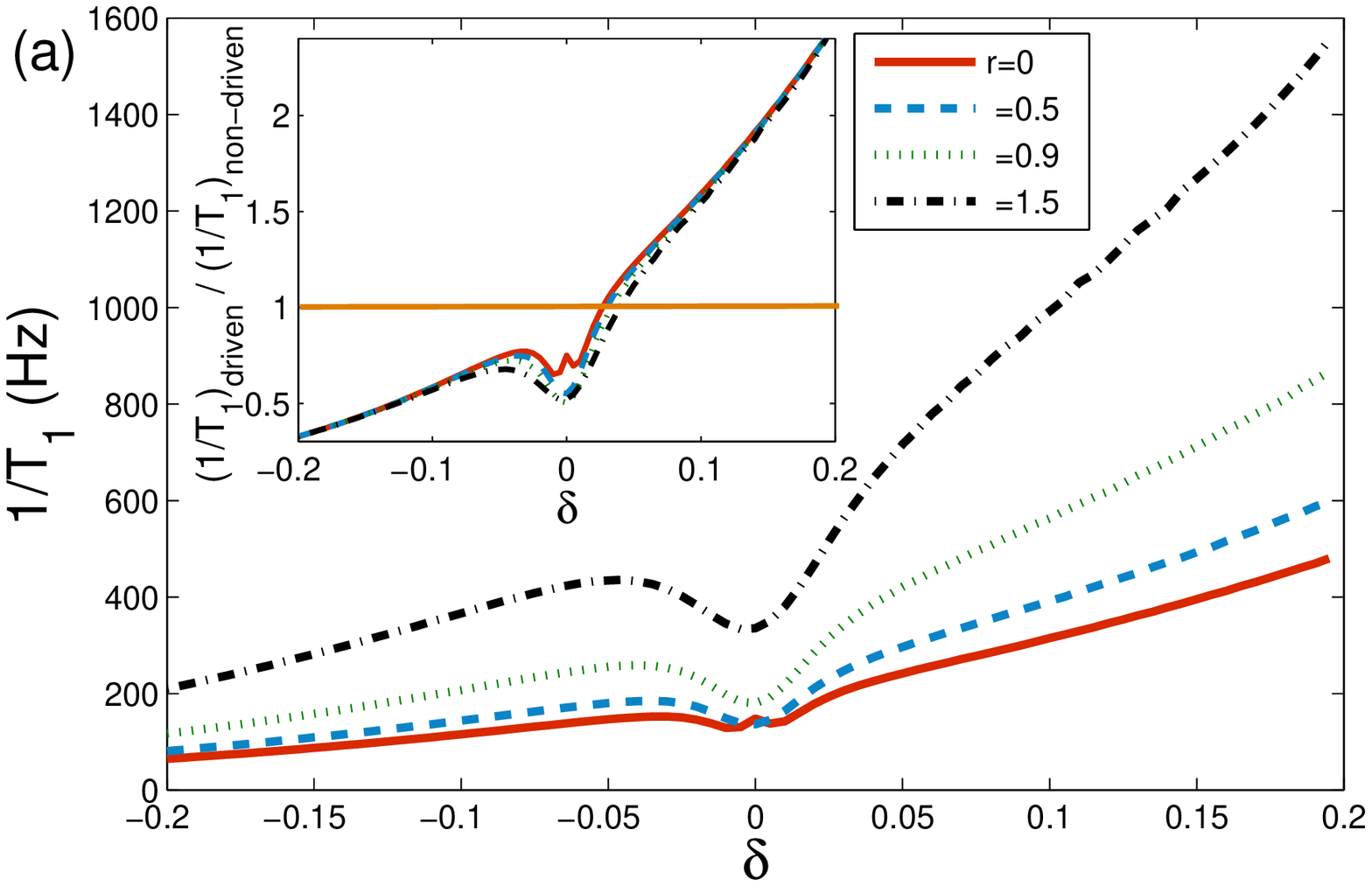}}
\subfigure{\label{T1nonab:phi}
\includegraphics[width=3.2in]{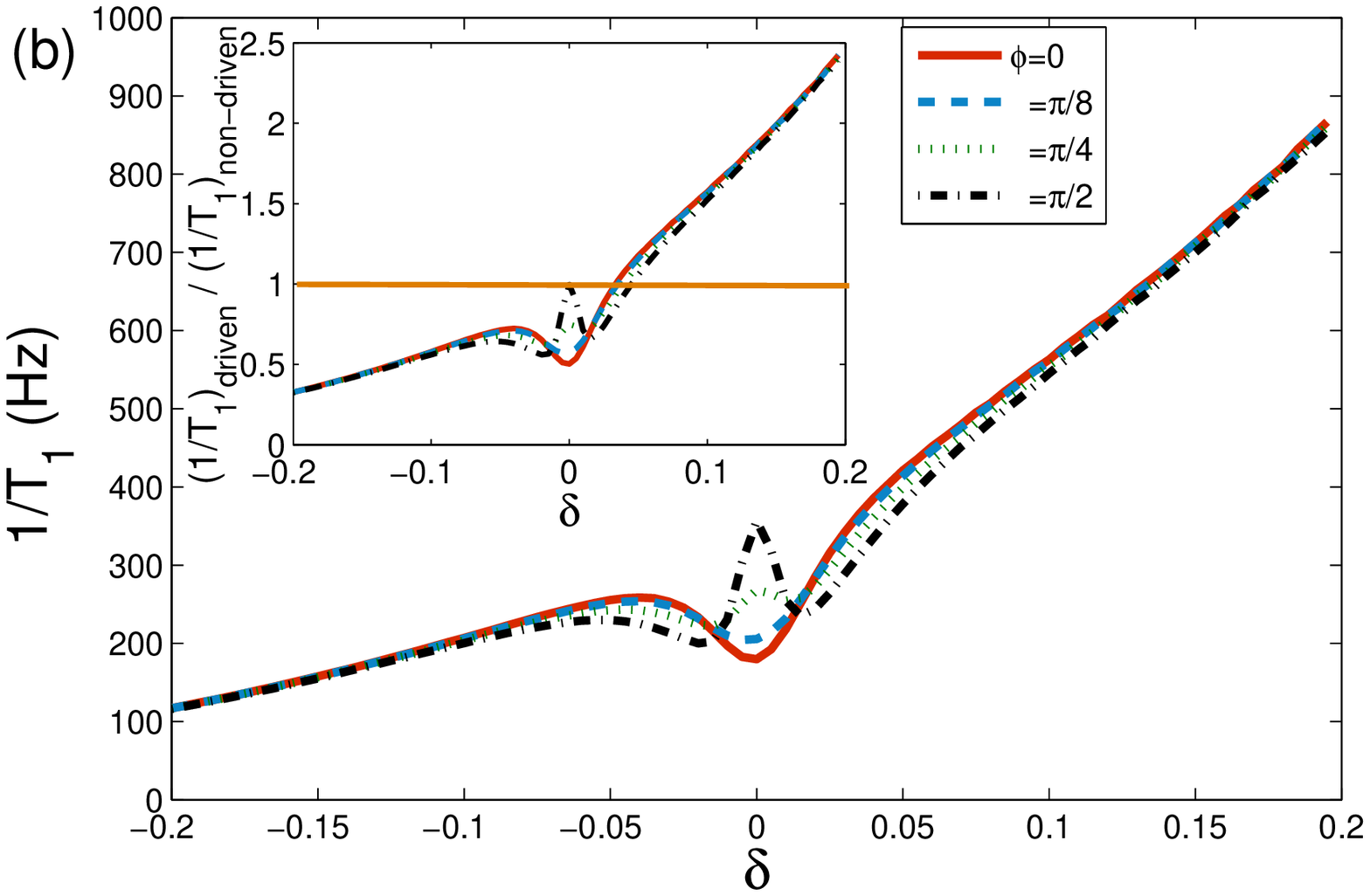}}
\caption{(Color online) Off-resonance relaxation rate $1/T_1$ of electron spin as a function of dimensionless frequency detuning $\de$ in presence of electrical noise for GaAs QD under the condition (a) different $r$ with $\phi=0$ and (b) different $\phi$ with $r=0.9$. We choose $\beta=1000$m/s, $B_z=1$T, $\hbar\om_d=1$meV, and $|E_x|=4000$V/m. Insets: the ratio $(1/T_1)|_{\rm driven}/(1/T_1)|_{\rm non-driven}$ vs. $\de$.}
\label{T1nonab}
\end{figure}

In Fig.~\ref{T1nonab} we plot $1/T_1$ as a function of the detuning $\de$, with various combinations of $r$ and $\phi$. There are several interesting features to the results. First, there is a general trend of increasing relaxation rate as the driving field frequency increases (and $\de$ increases from negative to positive values). Second, near resonance (small $|\delta|$), the relaxation rate has a strong $\phi$-dependence. Third, the relaxation rate has a strong $r$-dependence. Below we discuss these features in more detail.

At the large detuning limit, when $|\de| \gg R$, Eq.~(\ref{nonr}) indicates that $T_1 \rightarrow T'_1$. Numerically, the large-detuning limit is reached when $|\de| > 0.1 \gg R$, since $R \lesssim 0.01$ for EDSR. Physically, under the far off-resonance driving, the qubit is only slightly perturbed from its original state on the Bloch sphere, so that its relaxation rate in the lab frame should be close to that in the rotating frame. It is thus not surprising that $1/T_1$ shows the same increasing trend as $1/T_1'$, which is roughly proportional to $(1+\de)^5$ given by the phonon spectral density, and that $1/T_1$ is insensitive to $\phi$ at the large-detuning limit according to Eq.~(\ref{T1DeL}).

Near resonance, when $|\de| \ll R$, Eq.~(\ref{nonr}) shows that $1/T_1$ approaches $1/T'_2 = 1/2T_1' + 1/T'_\phi$. As we have shown in Fig. \ref{T1T2Tphi}, near resonance $1/T_1'$ has a dip, while $1/T'_\phi$ has a peak. The behavior of $1/T_1$ near resonance is thus a result of the competition between these two opposing contributions, and whether $1/T_1$ should have a peak or dip at resonance is mostly determined by the strength of $1/T'_\phi$, which in turn is strongly modified by the driving field phase shift $\phi$ when $r \sim 1$. Such a strong $\phi$-dependence by $1/T_1$ can be clearly seen in Fig.~\ref{T1nonab}(b) around $\delta = 0$.

Lastly, as indicated by Eqs.~(\ref{T1De}) and (\ref{TphiDe}), both $1/T_1'$ and $1/T'_\phi$ have a quadratic dependence on $r$. Since $1/T_1$ always falls between $1/T_1'$ and $1/T_2'$, $1/T_1$ should also have a quadratic dependence on $r$, as demonstrated in Fig.~\ref{T1nonab}(a). Interestingly, the inset of Fig.~\ref{T1nonab}(a) shows that the normalized relaxation rate $(1/T_1)|_{\rm driven}/(1/T_1)|_{\rm non-driven}$ is insensitive to $r$ at the large-$|\de|$ limit.  This is because $1/T_1'$ in Eq.~(\ref{T1DeL}) has the same $r$-dependence as $(1/T_1)|_{\rm non-driven}$ in Eq.~(\ref{nd}).

\subsection{Pure dephasing under off-resonance driving}\label{off_dephasing}

The transverse relaxation rate $1/T_2$ under the off-resonance driving field can only be obtained numerically in general. However, with special driving field phase shifts, the condition $\la\si_x(T_2)\ra=1/e$ could be expressed in a more compact and analytical form. For example, when $\phi=0$, using only the envelope functions from Eq.~(\ref{Six}), we have approximately
\begin{equation}\label{phinonr}
\frac{R\left(e^{-T_2/T_1'}-1\right)}{\sqrt{R^2+\de^2}}
+\frac{R^2 e^{-T_2/T_1'}}{R^2+\de^2}+\frac{\de^2 e^{-T_2/T_2'}}{R^2+\de^2}=\frac{1}{e},
\end{equation}
where $T_1'$ and $T_2'$ are the longitudinal and transverse relaxation rates in the rotating frame, determined by Eqs.~(\ref{T1De}) and (\ref{TphiDe}).

\begin{figure}[htbp]
\centering
\includegraphics[width=3.2in]{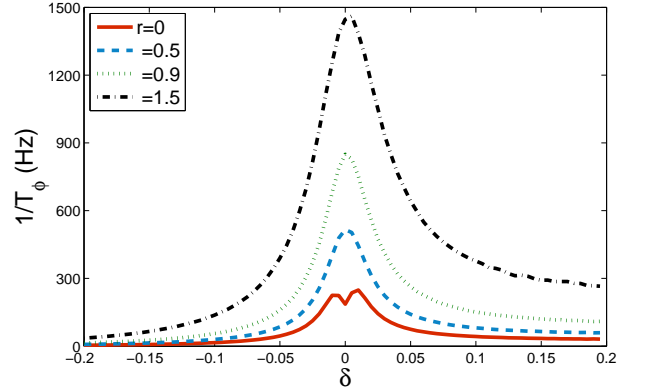}
\caption{(Color online) Off-resonance pure dephasing rate $1/T_\phi$ of electron spin as a function of dimensionless frequency detuning $\de$ in presence of electrical noise for GaAs QD under the condition with different $r$ with $\phi=0$.}\label{Tphi0non}
\end{figure}

The effect of detuning on the pure dephasing rate $1/T_\phi = 1/T_2 - 1/(2T_1)$ is demonstrated in Fig.~\ref{Tphi0non}. The $r$-dependence of the curves here has the same origin as in the case of relaxation rate $1/T_1$, while the peak structure of the curves in Fig.~\ref{Tphi0non} can be explained by examining the large- and small-detuning limits separately. At the large-detuning limit, when $|\delta| \gg  R$, Eq.~(\ref{phinonr}) shows that $T_2$ should approach $T_2'$. Since $T_1$ also approaches $T_1'$ in this limit according to Eq.~(\ref{nonr}), we obtain $1/T_\phi \approx 1/T_\phi' \approx 0$, as shown in Fig.~\ref{T1T2Tphi}. This suppression of pure dephasing is due to the fact that with off-resonance driving, the spin qubit maintains the original quantization axes; while phonon noise through EDSR leads only to transverse magnetic noise, so that it cannot cause pure dephasing. Near resonance, on the other hand, the driving field starts to cause Rabi flopping, so that qubit quantization axes are rotated and phonon noise could contribute to longitudinal magnetic fluctuations, and pure dephasing ensues. It is thus quite natural that all the curves in Fig.~\ref{Tphi0non} display a peaked structure around resonance.

\section{Discussions}\label{Sec:discussion}

Combining the relaxation and pure dephasing behaviors of a driven spin qubit, an attractive side-effect could emerge by exploiting the off-resonance driving.  Recall that single-qubit gates are often performed by a selectively resonant driving field, in the presence of other detuned qubits. As we discussed above, if the other qubits are negatively detuned (i.e. the selected qubit is positively detuned from the rest of the qubits), their overall decoherence rates would be slower than when they are not driven.

In the current study, we do not consider how driving could affect hyperfine-interaction-induced spin decoherence in a quantum dot, as we do not have a definitive semiclassical spectral density for the nuclear spin noise. Still, we could provide some qualitative assessment. In a finite magnetic field, the nuclear spin noise essentially causes only pure dephasing, with $S(\omega_Z) \ll S(0)$. In this case the slow but finite relaxation [mostly determined by $S(\Omega)$ where $\Omega \ll \omega_Z$ according to Eq.~(\ref{T_1})] for a driven spin qubit would still be faster than the negligible relaxation of a free spin qubit.  On the other hand, its pure dephasing is also determined by $S(\Omega)$ instead of $S(0)$ as indicated by Eq.~(\ref{phi0}), thus most probably it would be suppressed, and the overall decoherence represented by $1/T_2$ is slower. In essence, by driving the spin into a Rabi oscillation, we average out the effect of the nuclear spin noise and thus reduce the overall electron spin decoherence.

Our study of off-resonance driving also allows us to comment on the effect of the Overhauser field, which is a classical quasi-static mean field of the nuclear spin noise. The Overhauser field causes the spin qubit splitting to deviate from the Zeeman splitting due to the applied field. In a GaAs quantum dot, for example, this shift is up to $2$ mT. However, with an applied field that is larger than $0.1$ T, $\Delta/\omega_Z$ due to the Overhauser field is quite small, and should not affect the results obtained in this study, of spin decoherence due to phonon noise. The main effect of the Overhauser field is thus limited to gate errors when the driving field is used to generate gates.

\section{Conclusion}\label{Conc}

In conclusion, we have developed a general decoherence theory on a field-driven qubit. We find that driving, no matter resonant or off-resonance, can lead to significant modification of qubit relaxation and dephasing. In general, the qubit relaxation rate is determined not only by noise at its energy splitting $\omega_Z$, but also noise at sideband frequencies $\omega_Z \pm \Omega$, where $\Omega$ is the driving strength. In general, the changes in qubit decoherence depend on the driving frequency (or the detuning), driving strength, and the frequency-dependence of the spectral density of the environmental noise.  Our results could be relevant to decoherence control, general qubit manipulation, and sideband spectroscopy on a qubit, irrespective of the noise resource and its spectral function.

We have applied the general theory to the example of decoherence of a spin qubit in a semiconductor quantum dot driven through EDSR and in the presence of the electron-phonon interaction. We find that modifications to the spin decoherence rates depend closely on the ratio of Rashba/Dresselhaus spin-orbital coupling strength $r$, the phase shift $\phi$ of the elliptically polarized driving electric field, and the driving field detuning $\Delta$ or the dimensionless detuning $\delta$. In the near-resonance regime, the longitudinal relaxation rate is sensitive to the phase shift and is generally depressed by driving; while pure dephasing is often a more important decoherence mechanism than relaxation. In the far off-resonance regime, pure dephasing is strongly suppressed, while relaxation has a strong dependence on the frequency detuning of the driving field because of the $\omega^5$ spectral density of the phonon noise. In particular, when $\De<0$, i.e. $\nu<\om_Z$, the spin relaxation rate is suppressed relative to a non-driven spin.

\acknowledgments
We acknowledge financial support by US ARO (W911NF0910393) and NSF PIF (PHY-1104672). JJ also thanks support by NSFC grant No. 11175110.

\appendix

\section{Derivation of the EDSR effective Hamiltonian}\label{derivation}

In this appendix we derive the effective electron spin Hamiltonian when it is driven electrically via the spin-orbit (SO) interaction.

We consider an electron confined in a 2D quantum dot (QD) with finite confinement in the $x-y$ directions (the confinement along the growth direction $z$ is much stronger so that we neglect any orbital excitation along $z$).  The total effective-mass Hamiltonian for the electron consists of the kinetic energy $H_k$, the electric potential $V$ (which includes the 2D electrostatic confinement potential, the control field, and the electron-phonon interaction), the Zeeman splitting $H_Z$ caused by an applied magnetic field, and the SO coupling term $H_{\rm SO}$:
\begin{equation}
H_{\rm tot}=\frac{\ti{p}^2}{2m^*}+V[\vec{r}(t)] + \frac{1}{2}g\mu_B\vec{B}_0\cdot\vec{\si}+H_{\rm SO}.
\end{equation}
Here $\ti{p}=-i\hbar\nabla+\frac{e}{c}\vec{A}(\vec{r})$ ($e>0$), $m^*$ is the conduction electron effective mass ($0.067m_e$ in GaAs and $0.19m_e$ in Si, with $m_e$ the free electron mass), and $\mu_B\equiv\frac{e\hbar}{2m_e}\approx0.58\times10^{-4}$eV/T is the Bohr magneton.

In the absence of any driving field and noise, $V$ is a 2D electrostatic harmonic potential $V(\vec{r})=\frac{1}{2}m^*\om_d^2r^2$, where $\hbar\om_d\approx 1$meV is the confinement energy of the QD. When the electrical driving field and noise are introduced, in the form of an in-plane electric field $\vec{E}(t) = [E_x(t), E_y(t)]^T = [E_x^c+E_x^f, E_y^c+E_y^f]^T$ (with $\vec{E}^c$ being the control electric field and $\vec{E}^f$ the random field from whichever electrical noise), they cause a time-dependent displacement of the QD center, denoted by $\vec{r'}(t)$, so that the total potential becomes
\begin{equation}\label{VR}
V[\vec{r}(t)]=\frac{1}{2}m^*\om_d^2[\vec{r}-\vec{r'}(t)]^2, \quad \vec{r'}(t)=\frac{e\vec{E}(t)}{m^*\om_d^2}\,.
\end{equation}

With our choice of the coordinates ($x=[110]$, $y=[\bar{1}10]$ and $z=[001]$), the SO interaction term is expressed as:
\begin{equation}\label{HSO}
H_{\rm SO}=\beta_-\ti{p}_y\si_x+\beta_+\ti{p}_x\si_y,
\end{equation}
where $\beta_\pm \equiv \beta \pm \alpha$, and $\alpha$ and $\beta$ are the Rashba and Dresselhaus SO interaction strength. Notice that the form the SO interaction is closely related to the choice of the growth direction. For example, if the QD is in a heterostructure with a growth direction along $[111]$, the SO interaction would take on a different form.

To construct an effective spin Hamiltonian, we perform a unitary transformation~\cite{Golovach_PRL04, Fabian06, Huang} $e^SH_{\rm tot}e^{-S}$ on the total Hamiltonian, so that the electron spin and orbital degrees of freedom are decoupled to the first order of $H_{\rm SO}$.  The transformation matrix satisfies $[H_d+H_Z, S]=H_{\rm SO}$ with $H_d=\frac{\ti{p}^2}{2m^*}+V[\vec{r}(t)]$ and $H_Z=\frac{1}{2}g\mu_B\vec{B}_0\cdot\vec{\si}$. In other words, we are rotating the driving field and noise terms together with the system Hamiltonian, instead of treating them as perturbations. After the transformation, we obtain~\cite{Huang},
\begin{eqnarray}\label{Heff3}
H_{\rm eff}&\approx&\frac{1}{2}g\mu_B(\vec{B}_0+\vec{B}_e)\cdot\vec{\si}, \\
\non \vec{B}_e&=&\frac{2e}{g\mu_B\om_d^2}[\beta_-\dot{E}_y, \beta_+\dot{E}_x, 0]^T.
\end{eqnarray}
The driving electric field and the electrical noise now are transformed into an oscillating magnetic field and a magnetic noise. We recover Hamiltonian~(\ref{Heff}) in the general theory by requiring that the driving field and noise terms in Eq.~(\ref{Heff}) satisfy
\begin{eqnarray}\non
\frac{\Om}{2}\cos(\nu t+\phi) & = & \frac{e\dot{E}_y^c}{\hbar\om_d^2}\beta_-, \quad
n_x=\frac{e\dot{E}_y^f}{\hbar\om_d^2}\beta_-, \\
\frac{\Om}{2}\sin(\nu t+\phi) & = & \frac{e\dot{E}_x^c}{\hbar\om_d^2}\beta_+, \quad
n_y=\frac{e\dot{E}_x^f}{\hbar\om_d^2}\beta_+,
\end{eqnarray}
where $n_z=0$, and $\vec{B}_0=B_z\vec{z}$ with $g\mu_BB_z/\hbar = \om_Z$.  The driving electric field can thus be expressed as
\begin{eqnarray}\non
E_x^c & = & -\frac{\hbar\omega_d^2\Omega}{e\beta_+\nu} \cos(\nu t+\phi) \,, \\
E_y^c & = & \frac{\hbar\omega_d^2\Omega}{e\beta_-\nu} \sin(\nu t+\phi) \,,
\end{eqnarray}
which is elliptically polarized, with a ratio of $\beta_-/\beta_+$ for the magnitudes $E_x^c$ and $E_y^c$. In the special case when $\beta_- = 0$, we should go back to Eq.~(\ref{Heff3}), where $n_x=0$ and the effective field does not depend on $E_y$. We can thus fix $\phi=\pi/2$ and employ an electric field linearly polarized along the $x$-direction.  Equation~(\ref{Heff3}) would again take the form of Eq.~(\ref{Heff}).

In this study we have limited ourselves to an external field along the growth direction of the substrate. If the field has an in-plane component, the resulting driving field for the spin would contain an AC term in the longitudinal direction, making the problem much harder to be solved analytically by the current approach. We have made some numerical explorations in such cases, and our results so far do not show any remarkable difference from the results presented in the current manuscript.

\end{document}